\documentclass[sn-mathphys-num]{sn-jnl}

\usepackage{graphicx}%
\usepackage{multirow}%
\usepackage{amsmath,amssymb,amsfonts}%
\usepackage{amsthm}%
\usepackage{mathrsfs}%
\usepackage[title]{appendix}%
\usepackage{xcolor}%
\usepackage{textcomp}%
\usepackage{manyfoot}%
\usepackage{booktabs}%
\usepackage{algorithm}%
\usepackage{algorithmicx}%
\usepackage{algpseudocode}%
\usepackage{listings}%

\begin{document}

\title[A tissue-informed deep learning-based method for positron range correction in $^{68}$Ga PET imaging]{A tissue-informed deep learning-based method for positron range correction in preclinical $^{68}$Ga PET imaging}


\author*[1,2]{\fnm{Nerea} \sur{Encina-Baranda}}\email{nencina@ucm.es}

\author[1,2]{\fnm{Robert J.} \sur{Paneque-Yunta}}\email{rpaneque@ucm.es}

\author[1,2]{\fnm{Javier} \sur{Lopez-Rodriguez}}\email{javilo20@ucm.es}

\author[4,5,6]{\fnm{Edwin C.} \sur{Pratt}}\email{pratte@mskcc.org}

\author[4,5,6]{\fnm{Trong Nghia} \sur{Nguyen}}\email{nguyent8@mskcc.org}

\author[4,5,6]{\fnm{Jan} \sur{Grimm}}\email{grimmj@mskcc.org}

\author[3]{\fnm{Alejandro} \sur{Lopez-Montes}}\email{alejandro.lopez@unibe.ch}

\author[1, 2]{\fnm{Joaquin L.} \sur{Herraiz}}\email{jlopezhe@ucm.es}

\affil*[1]{\orgdiv{Nuclear Physics Group, EMFTEL and IPARCOS}, \orgname{University Complutense of Madrid (UCM)}, \orgaddress{\street{Av. Complutense, Pl. de las Ciencias, 1}, \city{Madrid}, \postcode{28040}, \state{Madrid}, \country{Spain}}}

\affil[2]{\orgdiv{Instituto de Investigación Sanitaria}, \orgname{Hospital Clínico San Carlos (IdISSC)}, \orgaddress{\street{Calle del Profesor Martín Lagos, s/n}, \city{Madrid}, \postcode{28040}, \state{Madrid}, \country{Spain}}}

\affil[3]{\orgdiv{Department of Nuclear Medicine}, \orgname{Bern University Hospital}, \orgaddress{\street{Freiburgstrasse 4}, \city{Bern}, \postcode{3010}, \country{Switzerland}}}

\affil[4]{\orgdiv{Molecular Pharmacology}, \orgname{Memorial Sloan Kettering Cancer Center}, \orgaddress{\street{1275 York Avenue}, \city{New York}, \postcode{10065}, \state{New York}, \country{USA}}}

\affil[5]{\orgdiv{Department of Radiology}, \orgname{Memorial Sloan Kettering Cancer Center}, \orgaddress{\street{1275 York Avenue}, \city{New York}, \postcode{10065}, \state{New York}, \country{USA}}}

\affil[6]{\orgdiv{Department of Pharmacology}, \orgname{Weill Cornell Graduate School}, \orgaddress{\street{1300 York Avenue}, \city{New York}, \postcode{10065}, \state{New York}, \country{USA}}}


\abstract{\textbf{Purpose:} Positron range (PR) limits spatial resolution and quantitative accuracy in PET imaging, particularly for high-energy positron-emitting radionuclides such as $^{68}$Ga. This study proposes a deep learning-based approach using 3D residual encoder-decoder convolutional neural networks (3D RED-CNNs), incorporating tissue-dependent anatomical information through a $\mu$-map-dependent loss function. We propose a model trained from realistic simulations and, using the information in the initial PET and CT images, obtain positron range corrected images. We validated our model in both simulations and real acquisitions.

\textbf{Methods:} Three different 3D RED-CNN architectures—Single-Channel, Two-Channel, and DualEncoder—were trained using simulated PET datasets and evaluated on both synthetic and real PET acquisitions from $^{68}$Ga-FH and $^{68}$Ga-PSMA-617 mouse studies. The performance of each model was compared to a standard Richardson-Lucy deconvolution (RL-PRC) approach using metrics such as mean absolute error (MAE), structural similarity index (SSIM), contrast recovery (CR), and contrast-to-noise ratio (CNR).

\textbf{Results:} In simulations, CNN-based methods achieved up to 19\% improvement in SSIM and 13\% reduction in MAE compared to RL-PRC. The Two-Channel model showed the highest CR and CNR values, recovering lung activity with 97\% agreement to the ground truth, compared to 77\% with RL-PRC. Noise remained stable for CNN models ($\sim$5.9\%), whereas RL-PRC increased noise by 5.8\%. In preclinical acquisitions, the Two-Channel model achieved the highest CNR in different tissues, while maintaining the lowest noise level (9.6\%). Although no ground truth was available for real data, analysis confirmed superior tumor delineation and reduced spillover artifacts with the Two-Channel model.

\textbf{Conclusion:} These findings demonstrate the potential of CNN-based PRC for improving quantitative PET imaging, particularly for $^{68}$Ga. CNN-based methods, particularly the Two-Channel model, outperformed conventional deconvolution in both simulated and real data. Future work will focus on enhancing model generalization through domain adaptation and hybrid training strategies, as well as extending the method to other high-energy PET radionuclides.}




\maketitle
\section{Background}\label{sec1}

The use of $^{68}$Ga in PET imaging is growing significantly due to its versatile applications in preclinical and clinical studies \cite{baum2012theranostics}, the availability of $^{68}$Ge/$^{68}$Ga generators \cite{roesch2012maturation}, and the approval of $^{68}$Ga-labeled tracers for clinical use \cite{hennrich202168ga}. This radionuclide is particularly advantageous for labeling both small compounds and macromolecules \cite{conti2016physics}, enabling a wide range of diagnostic uses, including cancer imaging with $^{68}$Ga-DOTATATE for neuroendocrine tumors \cite{mojtahedi2014value, haug2012role} and $^{68}$Ga-PSMA for prostate-specific membrane antigen (PSMA) targeting in prostate cancer \cite{bois202068ga, lenzo2018review, han2018impact}. Additionally, $^{68}$Ga is employed for infection and inflammation imaging \cite{xu2020research}, cardiac imaging for myocardial perfusion \cite{autio202068}, and bone imaging for detecting metastases in prostate cancer \cite{sachpekidis201868}. However, $^{68}$Ga emits positrons with an average large energy compared to other standard PET radionuclides such as $^{18}$F and $^{11}$C (see Table \ref{tab:pr}), resulting in one of the main limiting factors of spatial resolution and accurate quantification in PET \cite{levin1999calculation}. Accordingly, authors like L.M. Carter et al. \cite{carter2020impact} have characterized the positron range (PR) of standard and non-standard PET radionuclides to better assess their impact on image quality, which becomes more important as the intrinsic resolution of the state-of-the-art PET scanners improves. 

\begin{table}[h]
    \centering
    \caption{$\beta^+$ Yield, Energy, and Positron Range for $^{18}$F, $^{11}$C and $^{68}$Ga \cite{gonzalez2014positron}.}
    \begin{tabular}{llccccc}
        \toprule
        & & \multicolumn{2}{c}{$\beta^+$ energy (keV)} & \multicolumn{3}{c}{Positron Range (mm)} \\
        \cmidrule(r){3-4} \cmidrule(l){5-7}
        Isotope & $\beta^+$ Yield (\%) & Mean & End-point & Material & Mean & Max \\
        \midrule
        \textbf{$^{18}$F} & 96.7 & 249.3 & 633.5 & Lung & 1.85 & 7.49 \\
        & & & & Water & 0.57 & 2.16 \\
        & & & & Bone & 0.32 & 1.28 \\
        \midrule
        \textbf{$^{11}$C} & 99.8 & 385.6 & 960.2 & Lung & 3.35 & 12.4 \\
        & & & & Water & 1.02 & 3.67 \\
        & & & & Bone & 0.75 & 2.82 \\
        \midrule
        \textbf{$^{68}$Ga} & 88.0 & 836.0 & 1899.1 & Lung & 8.86 & 27.1 \\
        & & & & Water & 2.69 & 9.06 \\
        & & & & Bone & 1.44 & 4.89 \\
        \bottomrule
    \end{tabular}
    \label{tab:pr}
\end{table}

PR is defined as the distance traveled by the positrons from the emission to the annihilation points, resulting in anisotropic blurring in reconstructed PET images. PR is influenced by both the kinetic energy spectrum of the emitted positrons (which is isotope-dependent) and the electronic density of the tissue through which the positron travels before annihilation. Higher positron kinetic energies and lower tissue densities lead to larger PR \cite{cal2013positron}. The mean energy of the positrons emitted by $^{68}$Ga is 836.0 keV, significantly higher than that of the commonly used $^{18}$F (249.3 keV) \cite{gonzalez2014positron}. This energy difference results in a substantially increased PR for $^{68}$Ga, particularly evident in low-density tissues such as the lungs. As shown in Table \ref{tab:pr}, the mean PR in lung tissue increases from 1.85 mm for $^{18}$F to 8.86 mm for $^{68}$Ga, with the maximum PR expanding from 7.49 mm to 27.1 mm, respectively. This trend is also present in other tissues, such as water, which shows an increase from 0.57 mm to 2.69 mm (mean PR) and bone, from 0.32 mm to 1.44 mm. The pronounced effect highlights the need for positron range correction (PRC) when using high-energy positron emitters like $^{68}$Ga. PRC is essential to produce quantitative PET images with spatial resolution comparable to those obtained with $^{18}$F or $^{11}$C, ensuring accurate localization and quantification of radiotracer uptake, especially in heterogeneous environments.

Several approaches have been proposed to correct the blurring caused by PR within the reconstruction process \cite{gavriilidis2022positron}. Derenzo \cite{derenzo1986mathematical} proposed reducing PR effect in the projection domain by Fourier deconvolution combined with the Filtered Back Projection (FBP) reconstruction. However, this approach does not consider the heterogeneous nature of PR in different materials and increases noise through deconvolution, especially in narrow projection bins and for isotopes with high positron average energies \cite{derenzo1986mathematical}. Some PRC methods are based on modeling the Point Spread Function (PSF) within the System Response Matrix (SRM) used in iterative reconstruction \cite{cal2015tissue, cal2018improving, kertesz2022implementation}. While most reconstruction algorithms account for the PR of $^{18}$F in water by modeling a default PSF of the system, modeling the PR of non-standard PET isotopes, such as $^{68}$Ga, remains challenging because the PSF or SRM must be modified. Additionally, since PR depends on the specific local tissue properties, realistic PR models require the SRM to be evaluated at each acquisition, which is not computationally efficient \cite{lopez2020deep}. Other methods use material-dependent isotropic filters derived from CT or MRI anatomical images to deconvolve the PR effect with a combination of kernels \cite{cal2011study}.  Even though these methods have shown promising results, they still present limitations, especially at the boundaries of high-density regions such as bone tissue \cite{gavriilidis2022positron}. 

Recently, Gavriilidis et al. proposed a tissue-independent $^{68}$Ga-specific PRC using a convolution kernel modeling the PR of $^{68}$Ga in water using Monte Carlo simulations
 \cite{gavriilidis2024performance, gavriilidis2025impact}. This kernel is integrated with the OSEM and Q.Clear iterative reconstruction algorithms, incorporating point spread function (PSF) modeling and time-of-flight (ToF) information. Despite its potential, tissue-independent PRC can lead to inaccuracies, especially at tissue boundaries, due to over- or under-correction in different tissue types \cite{gavriilidis2024performance}. 

Vass and Reader proposed a method to address spatial resolution losses in reconstructed images caused by effects like PR \cite{vass2023synthesized}. This method involves forward projecting the reconstructed image using a user-defined virtual scanner geometry to generate synthetic sinogram data. The synthetic sinogram data are then used as input for image reconstruction, integrating a modeled PSF to correct PR-induced resolution degradation.

PRC can also be dealt with as a post-processing step by deconvolution of the reconstructed images. A simple and fast strategy is to simulate a kernel that characterizes the blur due to PR in a specific tissues (e.g. water, soft tissue, bone, lung, etc.) and perform a Richardson-Lucy deconvolution using the kernel to de-blur the reconstructed image \cite{rukiah2018investigation}. However, this method recovers the loss of spatial resolution without considering the anisotropic nature of PR in the boundaries of tissues and often increases image noise caused by the deconvolution process \cite{gavriilidis2022positron}.

With the popularization of artificial intelligence, methods using convolutional neural networks (CNNs) have been proposed to reduce the PR effect when working with $^{68}$Ga. Yang \cite{yang2021compensating} tested three different CNN models for PRC in preclinical PET imaging of $^{68}$Ga. The study used a preclinical PET scanner and 30 phantom models in Monte Carlo simulations, running each model twice: once with $^{68}$Ga (for CNN input images) and once with back-to-back 511-keV gamma rays. Although the recovery coefficient and spill-over ratio improved after correction, the evaluation was only done on phantom simulations and did not include information from density in the training, which may affect if it is applied to real acquisition. In previous works  \cite{lopez2020deep},  Herraiz et al. demonstrated the potential of employing a U-Net architecture \cite{ronneberger2015u} trained on simulated data to achieve fast and accurate PRC, incorporating anatomical information from the $\mu$-map during training. However, while the results showed improved performance when this information was included, no specific constraint was applied to ensure the proper use of the anatomical data.

In this work, we build on previous work \cite{lopez2020deep} including a novel network working on 3D patches instead of 2D slices and a tissue-dependent loss function to enforce consistency with the anatomic information provided by the $\mu$-map.  We also significantly expanded the training dataset, improving the model's ability to generalize across diverse scenarios. The use of multi-scale volumetric patches better describe the localized three-dimensional effect of PR when compared to 2D slices approaches. For the modified loss function, we explored three distinct strategies for integrating anatomical information from the $\mu$-map: indirectly (Single-Channel), as an additional input channel (Two-Channel), and through a dedicated encoder (DualEncoder). To benchmark performance, we compared our method with the standard Richardson-Lucy post-processing algorithm  \cite{rukiah2018investigation}. Notably, we also evaluated the accuracy and impact of our approach using $^{68}$Ga-FH and  $^{68}$Ga-PSMA-617 mice acquisitions, showcasing the impact of PRC on real studies.

\section{Materials and methods}\label{sec2}
\subsection{Synthetic PET data}\label{subsec1}
Computational animal models are commonly used to perform simulations with realistic materials and anatomies of animals in radiation transport and radiotracer biodistribution studies. In this work, MOBY \cite{segars2004development} and Digimouse \cite{dogdas2007digimouse} whole-body numerical phantoms were employed to simulate activity, material, and density distributions. The original Digimouse model was modified to introduce a variety of tumor shapes and tissues, such as the myocardium, prostate, and thyroid, which were not included in the original implementation. Unlike our previous work \cite{lopez2020deep}, which relied on PET/CT images from a single database, the use of different atlas provides increased control over tissue characterization and allows us to generate a variety of activity distributions within the body based on radiotracers used in PET imaging ($^{68}$Ga-PSMA, $^{68}$Ga-DOTATOC, $^{68}$Ga-DOTATATE, $^{18}$F-FDG), as well as additional synthetic distributions.

To simulate the PR effects, an adapted PenEasy \cite{sempau2006configuration} version was developed to generate positron annihilation distributions based on the initial positron emission distribution, using material and density maps from MOBY and Digimouse models. PenEasy tracks the path of each positron until its annihilation, taking into account their energy and all materials within the path using positron energy distributions simulated with PenNuc \cite{garcia2019pennuc}, which considers all potential decay branches and nuclear properties for a wide range of radionuclides. The formation of positronium was not considered and positrons were assumed to annihilate in the voxel where their energy dropped below 100 eV.

The positron annihilation distributions from the adapted PenEasy were introduced into MCGPU-PET \cite{herraiz2024mcgpu} to produce fast and realistic simulations based on the geometry of the Inveon PET/CT scanner \cite{constantinescu2009performance}. MCGPU-PET outputs were stored in sinograms with 147×168×1293 bins, maximum ring difference of 79, an axial compression factor (span) of 11, and a radial bin size of 0.795 mm. These sinograms were reconstructed using  GFIRST \cite{herraiz2011gpu}, an PSF-3D-OSEM algorithm implemented in CUDA using 60 iterations and 1 subset, to obtain 154×154×159 (x,y,z voxels) reconstructed images with a voxel size of 0.776×0.776×0.795 mm. Each reconstructed image was based on simulations of 2 million events, ensuring a realistic representation of PET acquisition conditions.

The attenuation maps were obtained from MOBY and Digimouse phantoms and converted to mm$^{-1}$ by defining three different materials and their respective reference densities \cite{berger_stopping-power_2017}: air (CT = 0, $\rho$ = 0.0012 g/cm$^{3}$), soft tissue (CT $\sim$110, $\rho$ = 1.0 g/cm$^{3}$), and bone (CT = 255, $\rho$ $\sim$1.85 g/cm$^{3}$). A piecewise linear mapping was applied: one segment interpolating from air to water, and another from water to bone, yielding a continuous density map across the CT volume. This map was then converted into a $\mu$-map by scaling the densities with linear attenuation coefficient for 511 keV gammas.

The workflow was implemented for both $^{18}$F and $^{68}$Ga, resulting in a database that includes two reconstructed PET images and the $\mu$-map image for each numerical phantom. We assume that the $^{18}$F PRC correction is included in the PSF of the system, which also includes other effects such as non-collinearity. Representative examples of the simulated, synthetic, and real PET datasets used in this study are shown in supplementary material. To improve reproducibility, 15 Digimouse-based numerical phantoms generated for this work have been uploaded to a Zenodo repository \cite{encina_baranda}. MOBY-based datasets are not included due to license restrictions, as well as real data acquisitions.

\begin{figure}[ht]
\centering
\includegraphics[width=1.0\textwidth]{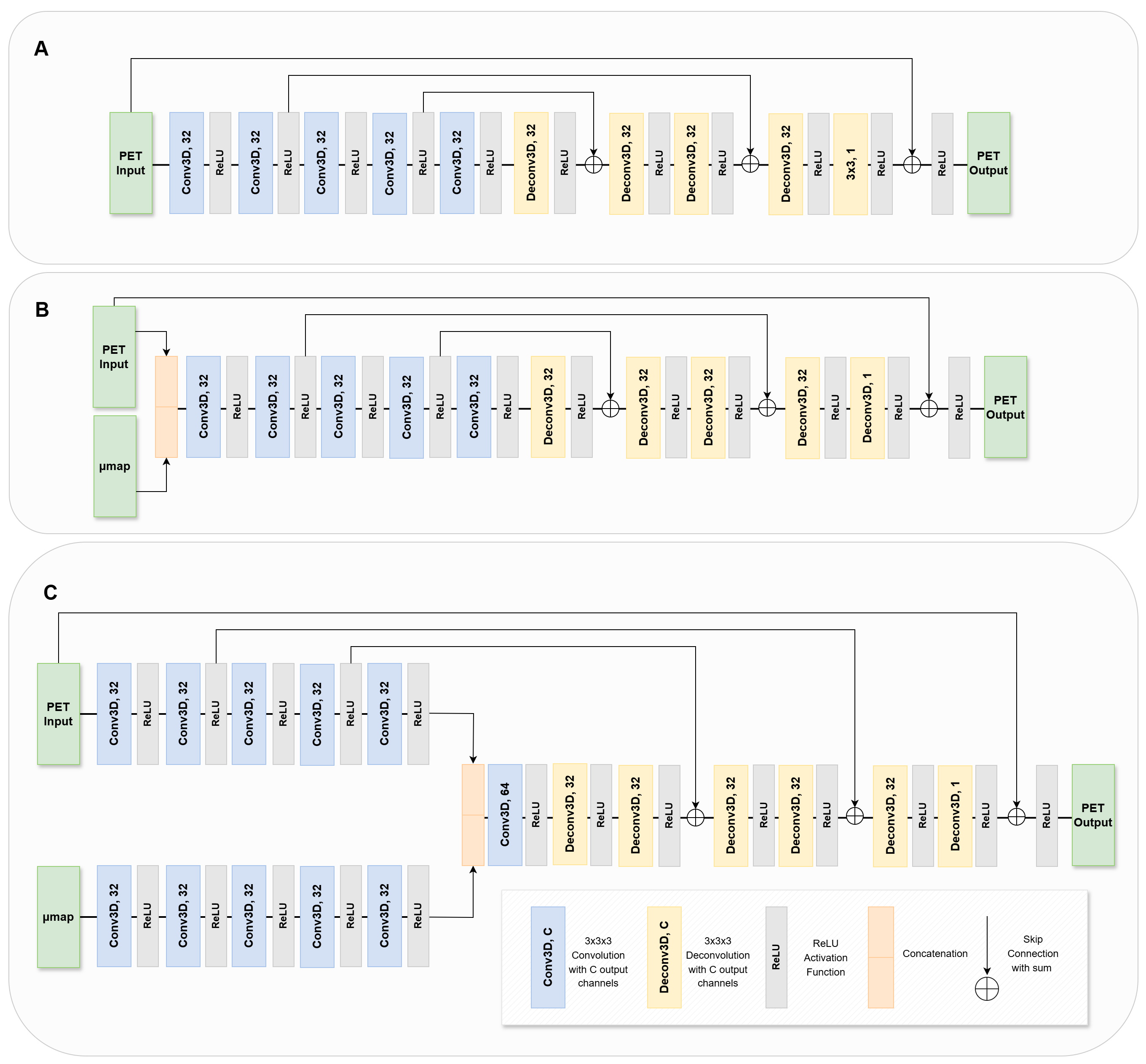}
\caption{3D RED-CNN architectures implemented: \textbf{A} shows the Single-Channel 3D RED-CNN, which processes only PET input; \textbf{B} represent the Two-Channel 3D RED-CNN, where $\mu$-map information is added as a second channel; and \textbf{C}, the DualEncoder 3D RED-CNN, where $\mu$-map  information is integrated into the network through a dedicated encoder. The legend in \textbf{C} indicates the network components: blue boxes represent 3×3×3 convolutional layers, orange boxes represent 3×3×3 deconvolutional layers, gray boxes denote ReLU activation functions, and connecting symbols skip connections with summation. The input images are previously normalized.}\label{networks}
\end{figure}
\subsection{CNNs models}\label{subsec2}
Three 3D residual encoder-decoder convolutional neural networks (3D RED-CNN) were developed. The model architectures are based on fully convolutional 2D CNN with shortcut connections between the encoder and the decoder proposed by Chen et al. \cite{chen2017low}, which has been demonstrated to outperform more complex models in low-dose CT denoising \cite{eulig2024benchmarking}.

Two key modifications were introduced to the original design: (a) 3D convolutional and deconvolutional layers replaced 2D layers to leverage 3D spatial information and mitigate artifacts between adjacent 2D slices \cite{chen2023total}; (b) the incorporation of anatomical information to enhance the network's performance. The integration of $\mu$-map  information was evaluated using three different architectures, as shown in Figure \ref{networks}. In the first NN architecture, the Single-Channel 3D RED-CNN processes only PET input, while the $\mu$-map  information is indirectly incorporated by influencing the loss function during training (Fig. \ref{networks}A). In the second NN, the Two-Channel 3D RED-CNN builds upon the Single-Channel architecture by adding an additional input channel, allowing direct incorporation of anatomical information from the $\mu$-map  alongside the PET input (Fig. \ref{networks}B). In the third architecture, the DualEncoder 3D RED-CNN processes the $\mu$-map  information through an additional encoder pathway, enabling a deeper integration of anatomical features into the network's feature extraction process (Fig. \ref{networks}C). These variations aim to explore the impact of different methods of incorporating anatomical information on the network's performance.

\subsection{Tissue-dependent loss function}\label{subsec3}
We use a loss function consisting of two components: (a) the Mean Absolute Error (MAE), a voxel-wise loss quantifying the error between the predicted image generated by the model and the ground truth PET image, and (b) a regularization term incorporating $\mu$-map information. To achieve this, Mutual Information (MI), a measure of the shared information between two random variables  $X$ and $Y$ \cite{cover1999elements}, can be used to force the NN to use the anatomical information. Formally, MI is defined as follows \cite{vergara2014review},
\begin{equation}
    \mathrm{MI}(X;Y) = \sum_{x\in X}\sum_{y \in Y} P(x,y) \cdot log \left( \frac{P(x,y)}{P(x) \cdot P(y)} \right) \
\end{equation}

where $P(x, y)$ is the joint probability distribution of $X$ and $Y$, and $P(x)\cdot P(y)$ is the product of the marginal distributions. When dealing with images, $X$ and $Y$ are sets of pixels $x$ and $y$, respectively. $\mathrm{MI}$ is commonly used in multimodal registration for PET/CT images \cite{maes1997multimodality}. 

Based on this concept, the differentiable global mutual information loss proposed by the open-source MONAI framework for DL in healthcare \cite{cardoso2022monai} was selected, which allows the implementation of MI as a loss function,  
\begin{equation}
    Loss(\hat{X};X;Y) = \mathrm{MAE}(\hat{X};X) + \lambda \cdot |\mathrm{MI}(X;Y)-\mathrm{MI}( \hat{X}; Y)|
\end{equation}

where $X$, $\hat{X}$ and $Y$ represents the target ($^{18}$F image), prediction ($^{68}$Ga PRC image) and  $\mu$-map, respectively. The weight factor $\lambda$ was set to 0.01, as it provided the best trade-off between preserving functional and anatomical consistency. The first term,  $\mathrm{MAE}(\hat{X};X)$, corresponds to the mean absolute error between the target and the prediction. The regularization is calculated as the absolute difference between the MI of the target and $\mu$-map and the MI of the prediction and $\mu$-map, which ensures that the $\mathrm{MI}(X;Y)$ of the output remains close to the expected value observed when using the target image. To ensure accurate computation of MI and minimize discrepancies from residual intensity scale differences—despite prior normalization—both PET and $\mu$-map patches were further standardized using Min-Max scaling during the calculation of the regularization loss.

\subsection{Implementation details}\label{subsec4}
All experiments were conducted on an NVIDIA GeForce RTX 2080 Ti (12GB VRAM) using CUDA 12.2 and implemented in PyTorch neural network framework. The dataset was divided into 17 numerical phantoms for training, 8 for validation, and 20 for testing. Instead of using full-resolution images during training, a volumetric patch-based sampling strategy was adopted to optimize memory usage and enhance the model’s ability to capture localized PR effects. Each epoch included 4200 patches sampled for training and 1800 patches sampled for validation. Patches of three different sizes (24×24×24, 18×18×18, 12×12×12 voxels) were randomly extracted to promote multi-scale learning and improve generalization \cite{moon2024m2former}. To prevent empty patches and improve robustness, a filtering step was applied to exclude empty regions based on the $\mu$-map. Given the physical constraints of PR effects, only random flipping in all three spatial directions (x, y, z) was applied. Scaling or zooming were avoided, as they could alter the PR effect, leading to inconsistencies in the network’s learning process.

\begin{figure}[ht]
\centering
\includegraphics[width=1.0\textwidth]{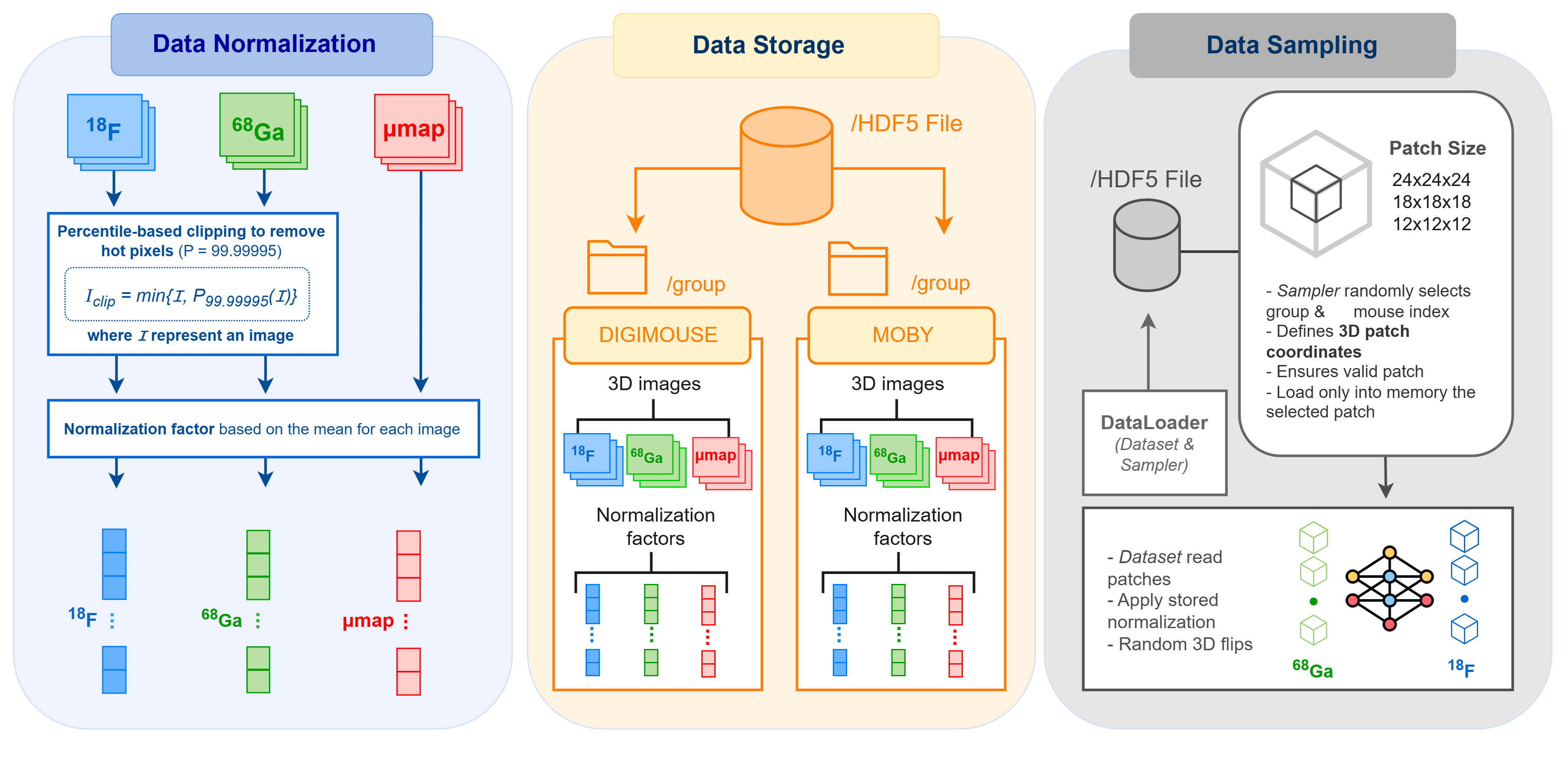}
\caption{Overview of the data preprocessing and sampling workflow. Each 3D image ($^{18}$F, $^{68}$Ga, and $\mu$-map) was first clipped at the 99.99995th percentile to remove hot pixels. A normalization factor was then computed for each volume and stored without modifying the raw data. All datasets and their corresponding normalization factors were saved in an HDF5 file, grouped by numerical phantom (Digimouse or MOBY). During training, the \textit{DataLoader} and custom \textit{Sampler} dynamically read patches from the HDF5 file, apply the stored normalization factor, and perform random 3D flips before feeding them into the network.}\label{implementation}
\end{figure}

The network was optimized using the AdamW optimizer \cite{loshchilov2017decoupled}, with an initial learning rate of $10^{-3}$, and a OneCycleLR scheduler \cite{smith2019super} dynamically adjusting learning rates and momentum to enhance training stability. The batch size was set to 120. Early stopping was applied with a patience of 50 epochs and a minimum required improvement of $10^{-7}$ in the validation loss.

To handle large volumetric PET datasets efficiently, the HDF5 file format \cite{neonhdf5} was used for data storage and sampling, allowing efficient random patch extraction without loading all the patches into memory, faster data access, and reduced I/O bottlenecks during training, as well as optimized GPU memory usage \cite{bieder2023diffusion}, enabling training on hardware with limited VRAM. Unlike some implementations that use mixed precision training to reduce memory usage, our approach was trained using full precision (FP32) to ensure numerical stability, preventing potential precision errors.

For each PET volume, voxel intensities above the 99.99995th percentile were clipped to avoid hot pixels due to reconstruction. Each PET and u-map image was then normalized by its mean intensity, yielding mean-equalized volumes with comparable intensity ranges across channels. The scaling coefficients were stored for later use during the training (see Figure \ref{implementation}) 

In this work, we leverage the properties of convolutional layers and their receptive fields to infer full images without patching \cite{luo2016understanding}. Convolutional operations learn spatial patterns locally and apply them uniformly across larger images due to shared weights and growing receptive fields. Additionally, spatial equivariance ensures that features learned in one region remain valid elsewhere \cite{cohen2019general}. By integrating volumetric patches during training and inferring full images during testing, our approach enhances efficiency while maintaining consistent feature extraction across the entire volume.

\subsection{Validation}
Validation was performed using simulated PET datasets, where both $^{68}$Ga and $^{18}$F PET images were generated under same conditions. The network was trained to predict the reconstructed $^{18}$F-equivalent distribution from the reconstructed $^{68}$Ga images while incorporating anatomical priors from the $\mu$-map. To assess correction accuracy, both global image comparisons and regional analyses were conducted to evaluate localized PR effects. The mean absolute error (MAE) and structural similarity index measure (SSIM) were used to quantify voxel-wise differences between corrected images and reference $^{18}$F PET images, ensuring that the PR correction restores radiotracer distribution while preserving structural integrity. All the results are given in Standardized Uptake Values relative (SUVr) to the average activity in the mouse. 

To further evaluate the impact of PR correction, contrast recovery (CR) and contrast-to-noise ratio (CNR) were analyzed in key anatomical regions, including the lungs, myocardium, bladder. For each tissue, two binary masks were used: one for the region of interest (ROI) and another for the background, matched in volume. CR was computed as the ratio of contrast levels between $^{68}$Ga-corrected and $^{18}$F reference images, ensuring proper recovery of activity distribution. CNR was calculated to assess whether PR correction maintained sufficient contrast while avoiding excessive noise amplification, following
\begin{equation}
    \mathrm{CNR} = \frac{\mathrm{Mean_{\text{ROI}}} - \mathrm{Mean_{\text{bckg}}}}{\mathrm{SD_{\text{bckg}}}}
    \label{eq3}
\end{equation}
where $\text{Mean}_{\text{ROI}}$ and $\text{Mean}_{\text{bckg}}$ denote the mean activity within the region of interest and the background, respectively, while $\text{SD}_{\text{bckg}}$ represents the standard deviation of the background activity.

Image noise was assessed using an spherical ROI in the liver region, assuming the uptake in this organ to be approximately uniform, making it a suitable reference for estimating statistical noise variations. Noise was quantified using the coefficient of variation, defined as the ratio of the standard deviation to the mean activity within a homogeneous ROI.

To establish a performance baseline, results from the proposed methods were compared to Richardson-Lucy Deconvolution (RL-PRC), a similar post-processing method in terms of time cost and ease of use. For this purpose, the $^{68}$Ga PSF was simulated in water using a point source centered in the FOV, following the same pipeline used for the numerical phantoms: PR simulated with PenEasy, followed by MCGPU-PET to generate the sinogram and GFIRST reconstruction to obtain a uniform PSF. The PSF was applied as an isotropic deconvolution kernel (8×8×8 voxels) to recover lost spatial resolution, using three iterations to balance resolution enhancement while preserving noise levels similar to the original PET images. The deconvolution process used the $^{68}$Ga PET image, the $^{68}$Ga PSF kernel, and a binary mask set to 1 within the mouse volume and 0 elsewhere to constrain the correction. The results from this post-processing approach were evaluated using the same criteria as those applied to the CNN-based methods.

\subsection{Preclinical $^{68}$Ga PET studies}\label{evalrealdata}
Experimental studies were conducted using the Inveon PET/CT system (Siemens Medical Solutions) at Memorial Sloan Kettering Cancer Center (New York, USA) to assess the feasibility of the proposed methods on the preclinical Inveon PET/CT scanner. The impact of the proposed PRCs were evaluated with two independent $^{68}$Ga acquisitions in mice. All PET images were reconstructed using GFIRST and corrected by random coincidences using a delayed time window approach within the acquisition \cite{brasse2005correction}, attenuation based on the values of the $\mu$-map and scatter using MCGPU-PET \cite{cabello2025comparison}.

The first acquisition involved a preclinical AKP TV model to induce pulmonary tumor metastasis, as the lung is the first site of colonization when the injection is done into the tail vein \cite{thies2020pathological}. To evaluate inflammation-associated tracer uptake, Feraheme (FH) nanoparticles were radiolabeled with $^{68}$Ga. FH is an iron oxide nanoparticle originally developed for the treatment of iron-deficiency anemia \cite{coyne2009ferumoxytol}, but it also functions as a tracer in PET imaging due to its selective uptake by monocytes cancer or inflammation \cite{yuan2018heat, gholami2020radio}. The mouse received an intravenous injection of 22.755 MBq of $^{68}$Ga-FH, and PET imaging was initiated 3 hours post-injection, with a total scan duration of 30 minutes. The energy window for this acquisition was set to 350–700 keV. As respiratory gating was not applied, the resulting images may exhibit motion blur from lung movement, making tumor identification and quantification challenging.

The second set of experiments involved a preclinical prostate cancer xenograft model where PSMA-positive PC3-PIP cells were implanted subcutaneously into the right shoulder of immunocompromised mice and allowed to develop tumors for four weeks prior to PSMA-targeted PET imaging \cite{boinapally2021prostate, mukherjee2016development, dam2017psma}. This model is widely used to study PSMA-specific therapies and diagnostics, as PC3-PIP cells express $\sim$2-fold higher PSMA levels compared to other engineered cell lines like LMD-PSMA \cite{mukherjee2016development}. The four-week growth period aligns with established protocols where tumor volumes typically reach measurable sizes \cite{boinapally2021prostate}. The mouse was injected with 1.93 MBq of $^{68}$Ga-PSMA-617, and PET scanning began 8 minutes post-injection, lasting 50 minutes. The energy window for this acquisition was set to 350-650 keV.

\section{Results}\label{sec3}
\subsection{Synthetic data analysis}
\begin{figure}[!h]
\centering
\includegraphics[width=0.99\textwidth]{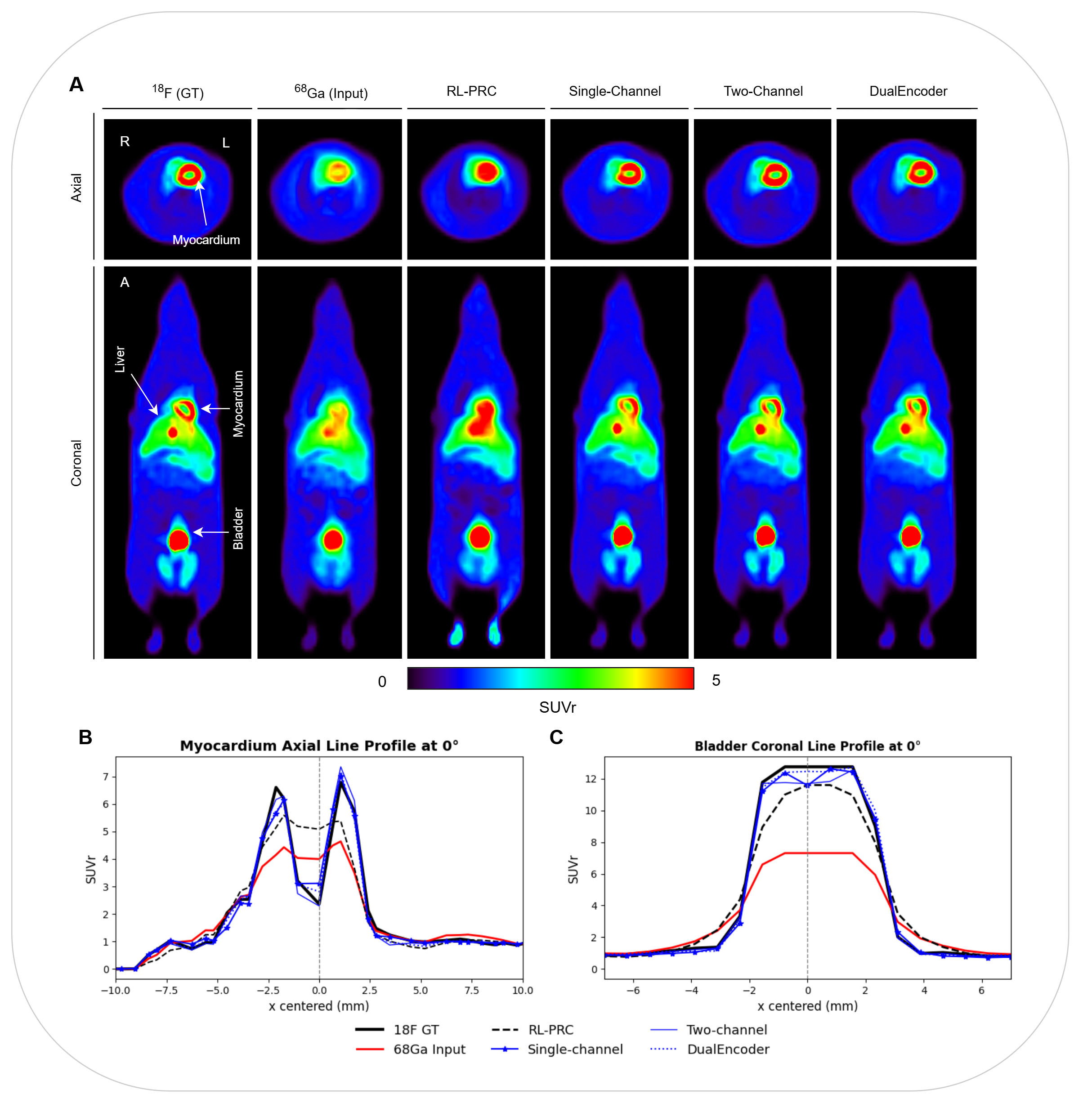}
\caption{\textbf{A} shows axial and coronal PET images comparing ground truth $^{18}$F with $^{68}$Ga input, $^{68}$Ga PET image PRC with Richardson-Lucy deconvolution and the output of the three models (Single-Channel, Two-Channel and DualEncoder 3D RED-CNN). The reconstructions aim to replicate the $^{18}$F ground truth distribution using $^{68}$Ga tracers through different correction methods. The white letters indicate the image direction: A=anterior, L=left, R=right. \textbf{B} shows the myocardium line profile in the axial view at 0$^{\circ}$ and \textbf{C} the bladder line profile in the coronal view at 0$^{\circ}$ . Abbreviations: GT=Ground Truth, RL-PRC=Richardson-Lucy Positron Range Correction. Results are given in Standardized Uptake Values relative (SUVr) to the mean activity in the mouse.}
\label{mcsimulation}
\end{figure}

 The Figure \ref{mcsimulation}A shows a comparison for one of the evaluated cases, displaying axial (top row) and coronal slices (bottom row) of the $^{18}$F image (defined as the ground truth, GT), the $^{68}$Ga image without PRC (i.e. with the PRC blurring), and the results from the Richardson-Lucy (RL-PRC) and the three 3D RED-CNN models. The uncorrected $^{68}$Ga image shows significant blur in both axial and coronal views compared to the $^{18}$F images, especially in areas of high activity such as the bladder and myocardium.
These results are supported by the line profiles along the myocardium and the bladder, shown in Fig \ref{mcsimulation} A and B, respectively. CNN-based approaches (blue curves) significantly improve SUVr compared to the $^{68}$Ga input and RL-PRC. Among them, the Two-Channel and DualEncoder models achieve the best alignment with the $^{18}$F image.  RL-PRC primarily restores activity but in the case of the bladder, predominantly recovers activity in the central region, resulting in inconsistent distribution across the tissue. The proposed methods maintain a uniform activity level throughout the organ, as observed in $^{18}$F.

\begin{table}[!ht]
    \centering
    \caption{Level of noise in liver (\%) from 20 mouse phantoms for ground truth, input, RL-PRC, and CNN-method images.}
    \begin{tabular}{cccccc}
        \toprule
        \multicolumn{6}{c}{\textbf{Level of Noise in Liver (\%)}} \\
        \multicolumn{6}{c}{Mean ± SD (N=20)} \\
        \midrule
        \textsuperscript{18}F (GT) & \textsuperscript{68}Ga (Input) & RL-PRC  & Single-Channel & Two-Channel & DualEncoder \\
        \midrule
        5.9 $\pm$ 1.1 & 7.8 $\pm$ 2.2 & 11.7 $\pm$ 5.9 & 5.8 $\pm$ 0.8 & 5.9 $\pm$ 0.9 & 5.9 $\pm$ 1.0 \\
        \bottomrule
    \end{tabular}
    \label{tab:noise}
\end{table}

\begin{figure}[ht]
\centering
\includegraphics[width=1.0\textwidth]{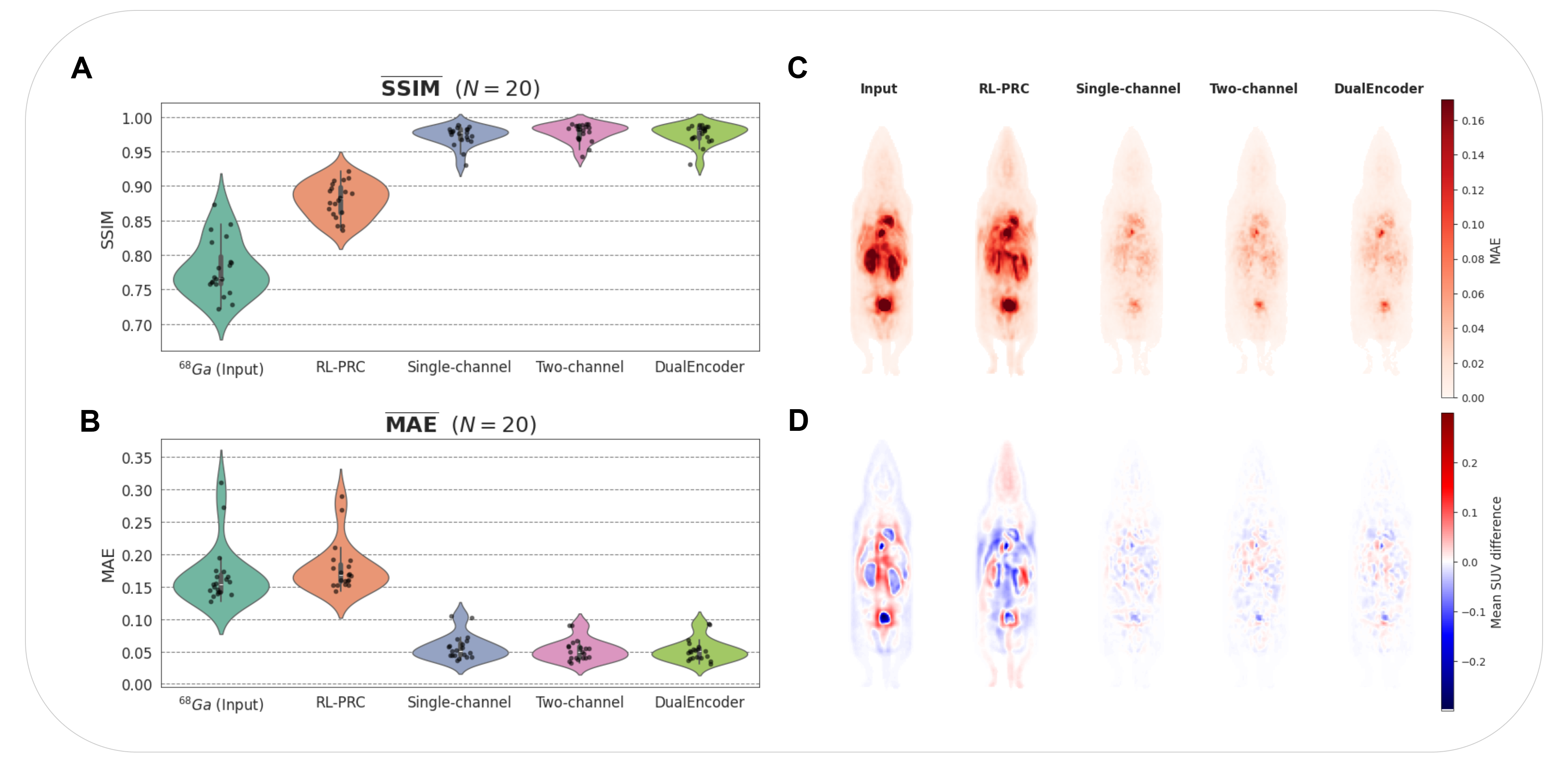}
\caption{Evaluation of image quality using SSIM and MAE for $^{68}$Ga images without and with PRC: \textbf{A} and \textbf{B} show violin plots displaying SSIM and MAE values for 20 samples. \textbf{C} and \textbf{D} show a slice coronal MAE and mean SUV difference maps of a single case. Abbreviations: RL-PRC=Richardson-Lucy Positron Range Correction, SSIM=Structural Similarity Index Measure, MAE=Mean Absolute Error. }
\label{violinplots}
\end{figure}

The SSIM and MAE values for 20 different simulated cases are shown in the violin plots in Figures \ref{violinplots}. The results show that all PRC methods improve upon the original $^{68}$Ga image. NN-based methods achieve 19$\%$  higher SSIM and 11$\%$ lower MAE than the uncorrected $^{68}$Ga image and outperform RL-PRC with approximately 9$\%$  higher SSIM and 13$\%$  lower MAE in all cases. Figure \ref{violinplots}C and D use MAE and mean SUVr difference maps to highlight regions where each PRC method struggles. For $^{68}$Ga, activity is underestimated in blue regions, while spillover is prominent around organs (red regions), due to the PR. RL-PRC recovers the activity from the mouse body to the tissues, while CNN-based methods are more consistent with respect to the ground truth. Noise levels measured in the liver are reported in Table \ref{tab:noise}. As anticipated, RL-PRC introduced a 5.8$\%$ increase in noise due to its iterative nature, while CNN-based methods maintained noise levels comparable to the GT images. 

\begin{figure}[h!]
\centering
\includegraphics[width=1.0\textwidth]{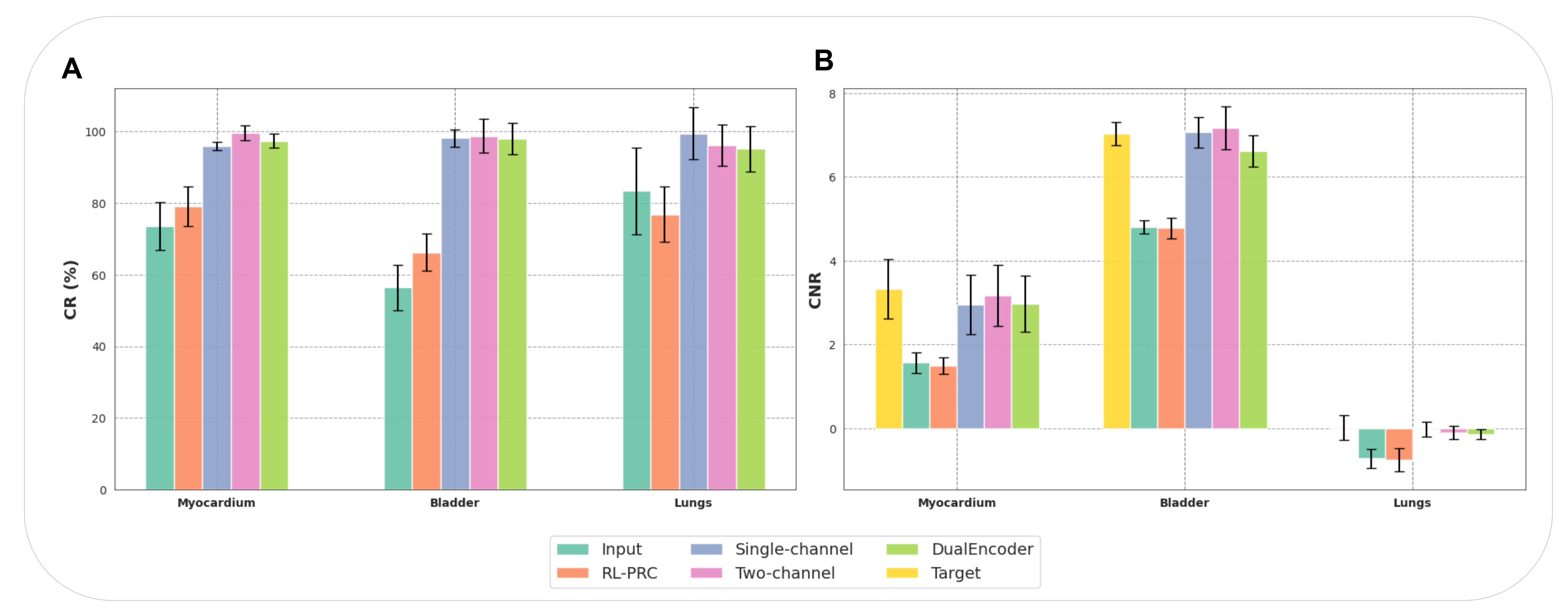}
\caption{\textbf{A} CR ($\%$) and \textbf{B} CNR comparison across organs -myocardium, bladder, lungs and liver- and methods.  Abbreviations: RL-PRC=Richardson-Lucy Positron Range Correction, CR=Contrast Recovery, CNR=Contrast-to-Noise Ratio,}
\label{barplots}
\end{figure}

In addition to the global image analysis, specific regions—myocardium, bladder and lungs—were also studied. Figure \ref{barplots} presents two bar charts summarizing the average CR and CNR values obtained from the 20 simulated samples. Across all tissues, CNN-based methods consistently outperform RL-PRC, demonstrating superior activity restoration and noise control. Particularly, in lungs RL-PRC leads to a reduction in CR, whereas the proposed CNN-based models recover activity more effectively. This difference is crucial since small-animal imaging is especially susceptible to PR effects, which can introduce vacuum artifacts in low-density regions, such as gases within the body and the lungs. In these regions, $^{68}$Ga exhibits a vacuum-like effect, characterized by activity underestimation within the lung and increased spillover at the boundaries. Since CNR is defined as equation \ref{eq3}, a less negative CNR indicates better correction performance by reducing spillover and improving contrast. All CNN-based models demonstrate notable improvements in CNR, effectively compensating for spillover at tissue interfaces while maintaining CR within the lungs, being the best performance from Single-Channel.

 \subsection{Preclinical $^{68}$Ga-FH and $^{68}$Ga-PSMA-617 analysis}
\begin{figure}[!ht]
\centering
\includegraphics[width=0.96\textwidth]{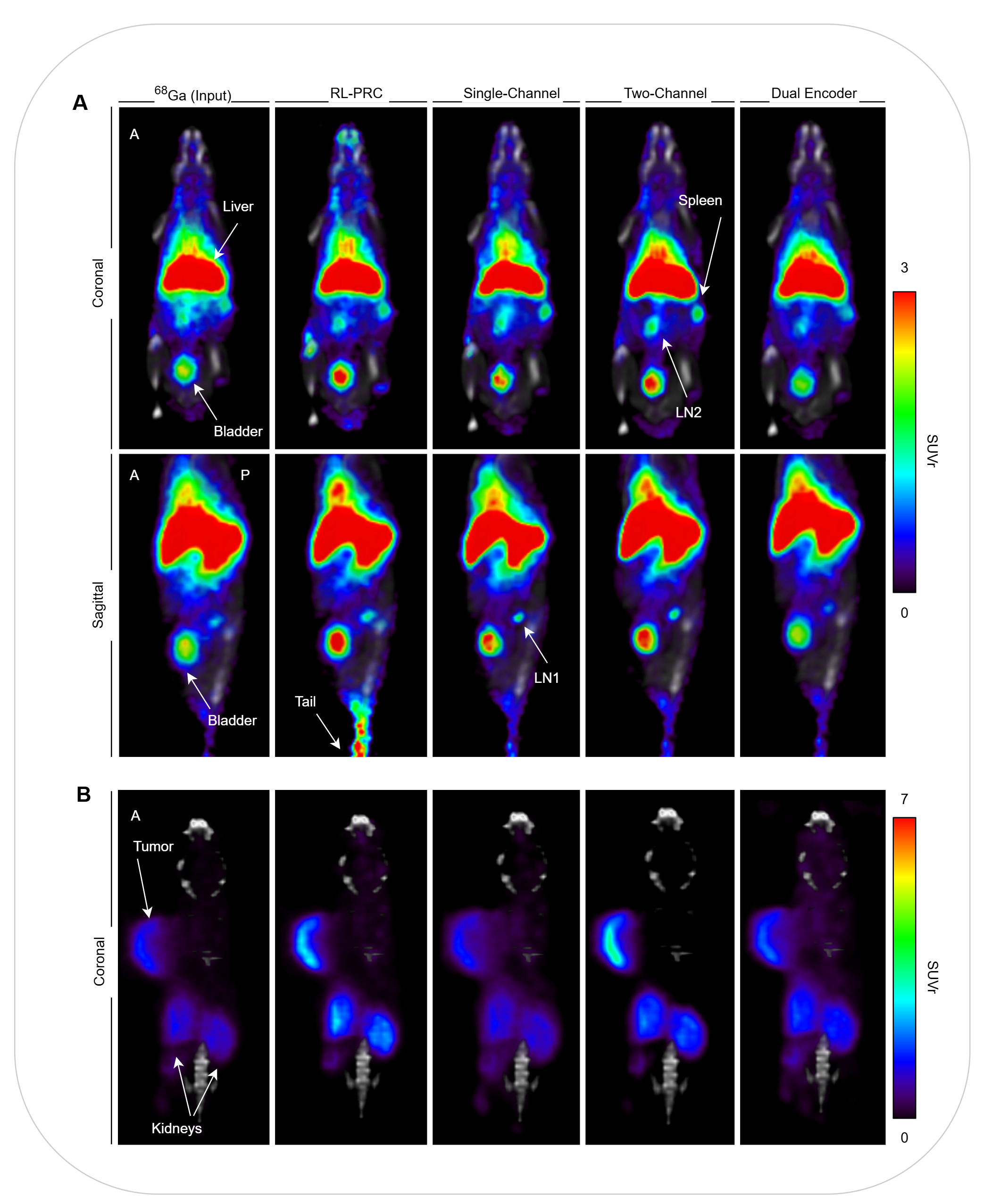}
\caption{Coronal views of PET images for $^{68}$Ga-FH \textbf{(A)} and $^{68}$Ga-PSMA-617 \textbf{(B)} acquisitions, comparing the uncorrected $^{68}$Ga input, Richardson-Lucy PRC (RL-PRC), and outputs from three DL models (Single-Channel, Two-Channel, and DualEncoder 3D RED-CNN). Additionally, a sagittal view is shown for $^{68}$Ga-FH. White letters indicate image orientation: A=anterior, P=posterior. Abbreviations: LN=Lymph Node. SUVr refers to the Standardized Uptake Value (SUV) ratio, calculated using the mean activity value in the mouse as a reference. The PET image is superimposed on the CT scan to help identify the location of tissues.} 
\label{mice1}
\end{figure}

The images reconstructed from real data before and after PRC were analyzed. In this case, no GT is avaliable. Before being input to the methods, PET images were preprocessed using the same pipeline as the simulations. Sinograms were formatted to match those generated by MCGPU-PET and reconstructed using GFIRST. All images were then normalized following the same protocol applied during training and testing, ensuring consistency across datasets. After inference, the number of counts in all output images was standardized to allow direct comparison across methods. Figure \ref{mice1} shows the uncorrected $^{68}$Ga input, alongside the outputs of the PRC methods for both $^{68}$Ga-FH and $^{68}$Ga-PSMA-617 acquisitions. Additionally, a sagittal view is included for $^{68}$Ga-FH.  It is important to note that the total number of counts is the same in the input (no-PRC) and the output (PRC) images.

In the $^{68}$Ga-FH images (Figure \ref{mice1}A), all PRC methods improved tissue definition, enhancing the visibility of the spleen and bowel in the coronal view. Among them, the Two-Channel model achieved the sharpest structural delineation, effectively reducing spillover artifacts around the liver, spleen, and lymph nodes (LN1 and LN2), and resulting in a more localized radiotracer distribution. In the sagittal view, RL-PRC introduced an artifact in the tail region, likely due to overcorrection. Notably, all methods demonstrated some degree of activity recovery in the lung region, with the Two-Channel model showing the most accurate performance. Additionally, the DualEncoder model effectively removes background but fails to effectively recover activity in the bladder. As summarized in Table \ref{tab:mice2}, the Single-Channel model achieved the highest CNR (38.2) in the LN1, followed closely by Two-Channel (37.0), RL-PRC (24.6) and significantly higher than DualEncoder (14.0). For both the bladder and spleen, the Two-Channel model demonstrated the best performance (49.0 and 22.0, respectively).

For $^{68}$Ga-PSMA-617, Figure \ref{mice1}B illustrates radiotracer uptake in both the xenografted tumor and kidneys. The results indicate that RL-PRC and the Two-Channel model achieved comparable tumor visualization; however, RL-PRC exhibited higher background activity, a pattern also observed in the Single-Channel and DualEncoder models. Table \ref{tab:mice2} shows a relative CNR increase of 11.5$\%$ for the Two-Channel model compared to the input, whereas the other methods resulted in a reduction of this metric.  This decrease is attributed to the use of the kidneys as the reference region for CNR calculation. While kidney uptake is similar to that of the tumor, it also presents a higher standard deviation, leading to a reduction in overall CNR values. Among all tested approaches, the Two-Channel model consistently achieved the highest CNR and lowest noise levels.

Regarding image noise, RL-PRC increased liver noise levels to 11.7$\%$, a trend also observed in DualEncoder (12.7$\%$). On the other hand, the Single-Channel (10.3$\%$) and Two-Channel (9.6$\%$) models slightly lower the lowest noise levels than the initial $^{68}$Ga image (11.7$\%$).

\begin{table}[!ht]
    \centering
    \caption{Comparison of CNR and noise levels for $^{68}$Ga-FH and $^{68}$Ga-PSMA-617 acquisitions across different PRC methods. Abbreviations: LN=Lymph Node.}
    \begin{tabular}{lccccc}
        \toprule
        & \multicolumn{4}{c}{\textbf{$^{68}$Ga-FH}} & \textbf{$^{68}$Ga-PSMA-617} \\
        \cmidrule(lr){2-5} \cmidrule(lr){6-6}
        & \multicolumn{3}{c}{CNR} & Noise (\%) & CNR \\
        \cmidrule(lr){2-4} \cmidrule(lr){5-5} \cmidrule(lr){6-6}
        & LN1 & Bladder & Spleen & Liver & Tumor \\
        \midrule
        $^{68}$Ga        & 13.6 & 23.7  & 11.8 & 11.7  & 2.42  \\
        RL-PRC           & 24.6 & 47.1  & 17.3 & 15.5  & 2.34  \\
        Single-Channel   & \textbf{38.2} & 47.7  & 21.2 & 10.3  & 2.13  \\
        Two-Channel      & 37.0 & \textbf{49.0}  & \textbf{22.0}  & \textbf{9.6}   & \textbf{2.70}  \\
        DualEncoder      & 14.0 & 26.2  & 11.5  & 12.7  & 1.98  \\
        \bottomrule
    \end{tabular}
    \label{tab:mice2}
\end{table}

\section{Discussion}

Positron range correction (PRC) is crucial for improving spatial resolution in PET imaging, particularly with radionuclides such as $^{68}$Ga. PR affects PET images, worsening the resolution and blurring the images. Furthermore, low density regions such as lungs, with lower probability of positron annihilation exhibit a loss of activity and spillover into peripheral areas. These artifacts distort radiotracer distribution, reducing quantitative accuracy and impacting image interpretation.

This study proposed and validated a DL-based PRC method using 3D RED-CNN architectures. The training framework incorporated PR in simulations with PenEasy \cite{sempau2006configuration} and MCGPU-PET \cite{herraiz2024mcgpu} that were then reconstructed, ensuring the model learned PR effects under realistic imaging conditions. We proposed a novel loss function able to integrate the anatomical information based on the Mutual Information (MI) between the PET and the $\mu$-map. Furthermore, we evaluated the integration of the $\mu$-map as an extra channel within the training of the models. Two-Channel strategy introduced the $\mu$-map at the beginning of the training process. DualEncoder explores two parallel encoder chains for the PET and the $\mu$-map merging them before the decoder step. Finally, Single-Channel did not include the $\mu$-map in the training process and the information of the $\mu$-map was only introduced in the tissue-dependent loss function. According to the results for simulated studies presented in Fig \ref{mcsimulation} and Fig \ref{violinplots}, the strategies including the $\mu$-map as an extra channel in the training outperformed the Single-Channel strategy, achieving an average SSIM 19.7\% and 19.4\% higher for Two-Channel and DualEnconder and respectively when compared to the PR corrupted input vs a 19.2\% improvement in the Single-Channel. All the studied strategies led to enhanced results compared to RL deconvolution where the average CR enhancement was of only 9.9\%. Moreover, as shown in Table \ref{tab:noise}, while the proposed methods based on DL successfully recovered the noise level of the target $^{18}$F images, RL deconvolution exhibited a noticeable increase (5.8\%) in the noise. This observation is consistent with the behavior observed by standard deconvolution methods based on iterative algorithms \cite{gavriilidis2022positron}.

The incorporation of $\mu$-map anatomical information into the loss function enables the model to learn tissue-dependent PR effects. This approach allows for regionally adaptive adjustments, enhancing activity recovery in low-density regions while mitigating blurring in high-activity soft tissues. Integrating physics-informed neural network (PINN) techniques into the modified tissue-dependent loss function ensures that the correction process is guided by the underlying physical properties of PET imaging, rather than relying solely on data-driven learning \cite{raissi2019physics, karniadakis2021physics, weikun2023physics}. By embedding tissue-dependent constraints, this approach improves quantitative accuracy, enhances structural consistency, and reduces the risk of artifacts, making it more robust to real PET acquisitions. These is particularly noticeable in the recovery of activity in low density regions such as lungs. Fig \ref{barplots} shows how using the CNN-based approaches, CR in lungs was recovered with a variation 97\% with respect to the ground truth for all of the studied networks, while conventional methods such as RL failed to recover the spillover in lungs with a variation of 77\% with the ground truth, very similar to the PR corrupted images (83\%). Also, while this study utilized $\mu$-maps from PET/CT, incorporating MRI-derived anatomical priors could further enhance PRC, particularly in PET/MR applications, where soft tissue contrast is superior. Future work should explore multi-modal strategies integrating MRI-based anatomical maps \cite{vandenberghe2015pet} to optimize PRC across different imaging modalities.

Both Digimouse and MOBY phantoms were used for training and testing to promote domain-invariant learning of PRC. Although both represent murine anatomy, they differ in structural detail and $\mu$-map origin: Digimouse $\mu$-maps are derived from real CT scans of an ex-vivo mouse, whereas MOBY $\mu$-maps are generated from an anatomical atlas with smoother, idealized tissue boundaries. To ensure compatibility, both datasets were harmonized to match the voxel size and simulation parameters of the Siemens Inveon PET system. This harmonization, while necessary, may have slightly reduced the structural detail of Digimouse and contributed to the minor residual differences observed, since more Digimouse cases were used for training, while only MOBY was used for evaluation. Nevertheless, the quantitative results (RMSE and SSIM) confirmed that the proposed method remains robust and generalizes well across domains.

The normalization strategy also plays an important role in ensuring stable and comparable learning across network architectures. The Single-Channel network, composed only of convolutional and ReLU layers, is inherently scale-independent. This property arises from the shift-invariant nature of convolutions \cite{lecun2015deep} and the positive homogeneity of ReLU activations, which allow the model to preserve proportional relationships in the input without requiring explicit normalization during inference. In contrast, the Two-Channel and DualEncoder architectures jointly process PET and $\mu$-map inputs, which differ in intensity distribution. Without normalization, these disparities could introduce bias in feature learning and affect convergence stability. By normalizing each input by its mean intensity, both modalities are brought to comparable scales, ensuring balanced contribution during training and consistent feature representation across channels.

Although MC–generated phantoms provide detailed anatomical and kinetic realism for training, they inevitably lack certain complexities of live small-animal PET data, such as respiratory and cardiac motion, detector non-uniformities, and unanticipated tracer heterogeneities. All CNN-based PRC methods outperformed RL-PRC when applied to simulated PET images, demonstrating superior contrast recovery, spatial preservation, and noise maintenance. However, their performance on real PET acquisitions faces the well-known challenge of generalizing models trained on synthetic data to real-world applications (simulation-to-reality gap) \cite{tremblay2018training, alkhalifah2022mlreal}. The Two-Channel architecture retained crisp tissue boundaries and minimal artifacts, whereas the DualEncoder model showed degraded performance. We attribute these failures to two interrelated factors: (1) over-reliance on $\mu$-map features can suppress tracer uptake, leading to reduced functional contrast, and (2) the DualEncoder’s greater representational capacity and complex feature pathways can cause it to memorize simulation-specific noise characteristics, leading to diminished generalization in vivo. To mitigate these issues, future work will improve the current training set is to use real $^{18}$F PET/CT acquisitions (instead of numerical phantoms) as ground truth and simulate the corresponding $^{18}$F and $^{68}$Ga PET images under identical conditions to provide a more complete and realistic training dataset. Strategies such as domain adaptation, transfer learning, and hybrid training approaches that integrate real and synthetic data have been explored in other fields and could provide potential solutions \cite{tobin2017domain, zhao2020sim}. 

For real $^{68}$Ga-FH and $^{68}$Ga-PSMA-617 acquisitions (Figure \ref{mice1}), accuracy was ensured by preserving the total number of detected counts, so that activity was only spatially redistributed. Compared with the Richardson–Lucy correction, the proposed method recovered physiologically plausible uptake in the tumor region without introducing spurious activity. Although ideal validation would involve paired acquisitions with isotopes with large and small PR, this is experimentally challenging because differences in tracer biodistribution may occur, even when the same tracer is used. Expert review by nuclear medicine physicians at MSKCC confirmed that the corrected images exhibited realistic uptake patterns, supporting the quantitative reliability of the proposed approach.

In this work, the MI was included in the loss function, including a regularization parameter to determine the strength of the MI term in the loss function, since MI captures complex, non-linear relationships beyond simple intensity correlations. To determine the optimal MI weight $\lambda$, we performed an ablation study on our simulated validation set, evaluating $\lambda$ = 0 (no MI), 0.001, 0.01, and 0.1. We found that 0.01 provided the best compromise: lower values ($\lambda$ $<$ 0.01) failed to enforce sufficient anatomical consistency, while higher values ($\lambda$ $\geq$ 0.1) resulted in “cross-talk” between PET and $\mu$-msp, manifesting as anatomical “hallucinations” that suppressed functional contrast and emphasized anatomical boundaries. By explicitly tuning $\lambda$, we ensure that the network leverages anatomical priors to guide the correction of positron-range blurring without overpowering the underlying radiotracer distribution. 

Working with volumetric patches during training allowed to capture the localized effect of PR while preventing the network from learning global activity distributions, focusing only on the correction of PR induced artifacts. Three different patch sizes (24×24×24, 18×18×18, and 12×12×12 voxels) were used to ensure the model learned PR effects at multiple spatial scales. The largest patch size allowed the network to capture the full extent of PR blurring in water ($\sim$ 9.2 mm for $^{68}$Ga), while smaller patches ensured learning of fine-scale features and high-frequency details in tissues. In cases involving different radionuclides, it is necessary to optimize and adapt the patch size to accurately reflect the radionuclide-specific PR characteristics.

At inference, the model processed the entire PET volume without requiring patch division, leveraging CNN properties such as shared weights and receptive fields. This approach enabled scalable inference on images of any size, as convolutional layers do not impose constraints on input dimensions, making the method adaptable to different FOV sizes. This implicit sliding window effect ensured seamless and consistent PRC across the entire image, eliminating potential stitching artifacts that could arise from patch-based inference.

Although this study primarily focused on 3D RED-CNN architectures, the proposed tissue-dependent loss function has the potential to be integrated into other DL models, such as U-Net \cite{ronneberger2015u}, Vision Transformers (ViT) \cite{dosovitskiy2020image, takahashi2024comparison}, and other modern NN architectures \cite{abdou2022literature}. Future research will explore alternative methods for incorporating $\mu$-map information, including direct network integration, the use of attention mechanisms for selective feature enhancement \cite{vaswani2017attention}, or multi-modal fusion techniques that combine PET and anatomical priors at different stages of the learning process \cite{huang2020review}. A comparative analysis of these approaches could further optimize PRC by identifying the most effective way to balance functional and anatomical information.

The findings of this work highlight the potential of DL-based PRC as a post-processing step for radionuclides with  large positron range effects. Further validation using diverse preclinical datasets is planned to assess model performance and establish a more realistic comparison with ground truth data. Additionally, extending the approach to other high-energy positron emitters, such as $^{82}$Rb and $^{124}$I, would broaden its applicability, contributing to improved quantitative accuracy in PET imaging across multiple research settings.

\section{Conclusion}
This study presents a tissue-informed deep learning approach for PRC in $^{68}$Ga PET imaging, integrating anatomical priors into a 3D RED-CNN framework. By incorporating tissue-dependent information through a $\mu$-map-guided loss function, the proposed method enables regionally adaptive PRC, mitigating spillover in low-density tissues and reducing blurring in high-activity regions.

Quantitative evaluation demonstrated that all CNN-based PRC models outperformed RL-PRC in simulations, with the Two-Channel model achieving the highest CR, CNR, and same noise levels. Similarly when applied to real PET data, the Two-Channel approach remained the most robust, effectively preserving tumor delineation in $^{68}$Ga-PSMA-617 imaging while reducing noise. However, more studies are needed to further validate this PRC with real data, specially looking for references for quantitative assessment. 

\backmatter
\section*{Declarations}
\begin{itemize}
\item Funding - We acknowledge support from the following grants: FASCINA (PID2021-126998OB-I00), Prototwin Project (TED2021-349130592B-I00), and 3PET (PDC2022-133057-I00), funded by the Spanish Ministry of Science, Innovation and Universities (MCIU) through AEI/10.13039/501100011033 and the European Union’s NextGenerationEU/PRT program. This work received support within the framework of the LUNABRAIN-CM (TEC-2024/TEC-43) project, funded by Comunidad de Madrid (Spain) through the R$\&$D activities program in technologies, granted by Order 5696/2024. We also acknowledge funding from the TARTAGLIA Project (MIA.2021.M02.0005), supported by the Spanish Ministry of Economic Affairs and Digital Transformation as part of the Recovery, Resilience, and Transformation Plan, financed through the European Union’s NextGenerationEU funds. Additional support was provided by the National Institutes of Health (NIH), USA, under grants R01EB033000, 1K99CA276804-01A1, and the Cancer Center Core Grant P30-CA008748. We also acknowledge funding from the TAU Project (PR47/21 TAU-CM), supported by the NextGenerationEU/PRTR funds.

\item Conflict of interest/Competing interests - No conflict of interest 
\item Ethics approval and consent to participate - All animal studies were conducted with the approval and adherence to the MSKCC IACUC under protocol 08-07-014.
\item Consent for publication - Not applicable
\item Availability of data and materials - The datasets generated and/or analysed during the current study are available in the Zenodo repository: \url{https://doi.org/10.5281/zenodo.15271216}
\item Author contribution - Monte Carlo simulations, neural network training and analysis of results was performed by Nerea Encina-Baranda. Neural network search and efficient data storage was prepared by Javier Lopez-Rodriguez. Help with the Monte Carlo simulations to include the positron range was carried out by Robert J. Paneque-Yunta. Acquisition of preclinical cases was carried out by Edwin C. Pratt, Trong Nghia Nguyen and Jan Grimm. Design and guidance of the project was carried out by Joaquin L. Herraiz and Alejandro Lopez-Montes. The first draft of the manuscript was written by Nerea Encina-Baranda. All authors have read and agreed to the final manuscript.
\item Acknowledgements - 

\end{itemize}

\noindent
\bibliography{ref}

\end{document}